\documentclass[conference]{IEEEtran}
\IEEEoverridecommandlockouts

\usepackage{xcolor,soul,framed} 

\colorlet{shadecolor}{yellow}
\usepackage[pdftex]{graphicx}
\graphicspath{{../pdf/}{../jpeg/}}
\DeclareGraphicsExtensions{.pdf,.jpeg,.png}

\usepackage[cmex10]{amsmath}
\usepackage{array}
\usepackage{mdwmath}
\usepackage{mdwtab}
\usepackage{eqparbox}
\usepackage{url}

\usepackage{amssymb}
\graphicspath{ {./images/} }
\usepackage{orcidlink}
\usepackage{breqn}
\usepackage{physics}

\usepackage{amsmath}

\usepackage{tikz} 

\usepackage{float}

\usepackage{algorithm} 
\usepackage{algpseudocode}

\usepackage{breqn}
\usepackage{physics}
\usepackage{subcaption}

 \usepackage{adjustbox}

\usepackage{booktabs}  
\usepackage{array}     
\usepackage{multirow}  
\usepackage{graphicx}  

\usepackage{blindtext}
\usepackage{xcolor}
\usepackage{hyperref}
\usepackage{graphicx}
\graphicspath{ {./images/} }
\usepackage{diagbox}
\usepackage{amsmath} 
\usepackage{mathtools}
\usepackage{lipsum}
\usepackage[nocompress,space]{cite}
\usepackage{amssymb}
\usepackage{pdfrender}
\usepackage{blindtext}
\usepackage{xcolor}
\usepackage{hyperref}
\usepackage{graphicx}
\graphicspath{ {./images/} }
\usepackage{diagbox}
\usepackage{amsmath}
\usepackage{mathtools}
\usepackage{lipsum}
\usepackage[nocompress,space]{cite}
\usepackage{pdfrender}
\usepackage[
  separate-uncertainty = true,
  multi-part-units = repeat
]{siunitx}
\usepackage{textcomp}
\usepackage{siunitx}
\usepackage{float}
\usepackage{caption}
\usepackage{adjustbox}
\usepackage{rotating}
\usepackage{blindtext}
\usepackage{xcolor}
\usepackage{hyperref}
\usepackage{graphicx}
\graphicspath{ {./images/} }
\usepackage{diagbox}
\usepackage{amsmath}
\usepackage{mathtools}
\usepackage{lipsum}
\usepackage[nocompress,space]{cite}
\usepackage{pdfrender}
\usepackage{soul,color}
\usepackage{orcidlink}

\usepackage{xcolor,soul,framed} 

\colorlet{shadecolor}{yellow}
\usepackage[pdftex]{graphicx}
\graphicspath{{../pdf/}{../jpeg/}}
\DeclareGraphicsExtensions{.pdf,.jpeg,.png}

\usepackage[cmex10]{amsmath}
\usepackage{array}
\usepackage{mdwmath}
\usepackage{mdwtab}
\usepackage{eqparbox}
\usepackage{url}

\usepackage{amssymb}
\graphicspath{ {./images/} }
\usepackage{orcidlink}
\usepackage{breqn}
\usepackage{physics}

\usepackage{tikz} 

\usepackage{float}

\usepackage{multirow}
\usepackage{diagbox}

\usepackage{graphicx}  

\usepackage[T1]{fontenc}

\usepackage{threeparttable}  

\usepackage{pifont}

\usepackage{bbding}
\usepackage{wasysym}

\usepackage{cleveref}

\usepackage{makecell} 

\newif\ifshowmods
\showmodstrue     

\ifshowmods

\else

\fi

 \makeatletter
\newcommand*\bigcdot{\mathpalette\bigcdot@{.5}}
\newcommand*\bigcdot@[2]{\mathbin{\vcenter{\hbox{\scalebox{#2}{$\m@th#1\bullet$}}}}}
\makeatother

\usepackage{longtable}

\begin{document}

\title{

Machine Learning-Based Cyberattack Detection and Identification for Automatic Generation Control Systems Considering Nonlinearities

}

\author{\IEEEauthorblockN{
Nour M. Shabar\orcidlink{0009-0006-2935-2263}\IEEEauthorrefmark{1}, Ahmad Mohammad Saber\orcidlink{0000-0003-3115-2384}\IEEEauthorrefmark{6},
and
Deepa~Kundur\orcidlink{0000-0001-5999-1847}\IEEEauthorrefmark{6} 
\IEEEauthorblockA{\IEEEauthorrefmark{1}Department of Electrical Engineering, Khalifa University, Abu Dhabi, United Arab Emirates}
\IEEEauthorblockA{\IEEEauthorrefmark{6}Department of Electrical and Computer Engineering, University of Toronto, Toronto, ON, Canada}
}
}

\maketitle

\begin{abstract}
Automatic generation control (AGC) systems play a crucial role in maintaining system frequency across power grids. However, AGC systems' reliance on communicated measurements exposes them to false data injection attacks (FDIAs), which can compromise the overall system stability. This paper proposes a machine learning (ML)-based detection framework that identifies FDIAs and determines the compromised measurements. The approach utilizes an ML model trained offline to accurately detect attacks and classify the manipulated signals based on a comprehensive set of statistical and time-series features extracted from AGC measurements before and after disturbances. 
For the proposed approach, we compare the performance of several powerful ML algorithms. 
Our results demonstrate the efficacy of the proposed method in detecting FDIAs while maintaining a low false alarm rate,
with an F1-score of up to 99.98\%, outperforming existing approaches. 

\end{abstract}


\section{Introduction}

Large power systems consist of multiple interconnected areas, each with its generation units and loads. A centralized control system, known as Automatic Generation Control (AGC), is responsible for maintaining the stability and frequency of the power system \cite{singh2017distributed}. The AGC system calculates the Area Control Error (ACE) for each area, which determines the necessary corrective actions to maintain nominal frequency. Based on the ACE, the required generation for each area is adjusted accordingly \cite{variani2013distributed}. The AGC system relies on frequency and tie-line power measurements to compute ACE. Protecting the AGC system from cyberattacks is critical, as such attacks can lead to system instability and significant frequency deviations \cite{khalaf2017detection}. Additionally, the AGC system exhibits important nonlinearities, including Governor Dead-Band (GDB), Generation Rate Constraints (GRC), and communication time delays \cite{golpira2011application}, which affect its response to disturbances and attacks.

False data injection attacks (FDIAs) pose a significant threat by manipulating AGC measurements \cite{choeum2021trilevel}. While AGC systems incorporate bad data detection mechanisms to filter out large anomalies, sophisticated FDIAs are designed to bypass these mechanisms and remain undetected \cite{khalaf2019joint}. Several detection strategies have been proposed to counteract FDIAs on AGC systems. In \cite{ameli2018attack}, an observer-based approach was introduced to estimate frequency deviations and classify attacks. Similarly, an online detection model leveraging dynamic watermarking was developed in \cite{huang2018online} to detect manipulated AGC signals. 
A different approach to signal watermarking was developed in \cite{Roy2024online} by transforming time series measurements into unique watermarked images. In \cite{Mohammadi2024online}, an intrusion detection approach has been proposed for protecting AGC systems.
However, most previous works overlook the impact of AGC system nonlinearities, limiting the practicality of their detection methods. Recent studies emphasize the importance of accounting for these nonlinearities—GDB, GRC, and transportation time delay—to improve detection accuracy \cite{musleh2023attack,ayad2022mitigation,roy2022machine}. In \cite{roy2022machine}, a multi-agent model was integrated with AGC for attack detection, but it only considered GRC. A data-driven approach using Long Short-Term Memory (LSTM) autoencoders was proposed in \cite{musleh2023attack} to detect FDIAs while considering GDB and GRC. More recently, Ayad et al. \cite{ayad2022mitigation} developed an LSTM-based classifier trained directly on AGC measurements to detect FDIAs while incorporating all three sources of nonlinearity. However, the black-box nature of their approach limits its interpretability, making it challenging for real-world adoption.

In this regard, this paper aims to contribute by developing a machine learning (ML)-based framework for detecting and classifying FDIAs in nonlinear AGC systems. Our approach utilizes an ML model trained on extracted features from AGC measurements, enabling the effective identification of manipulated signals. By leveraging feature-based learning with powerful ML models of interpretable performance rather than end-to-end deep learning, our method enhances transparency, allowing human power system operators to understand the features influencing detection decisions. Our results demonstrate that the proposed approach achieves high detection accuracy with a low false alarm rate, outperforming existing FDIA detection schemes.

\section{AGC System Model, Nonlinearities and Vulnerability to FDIAs}\label{Section:Preliminaries}

\begin{figure}[t!]
    \centering
    \includegraphics[width=1\columnwidth]{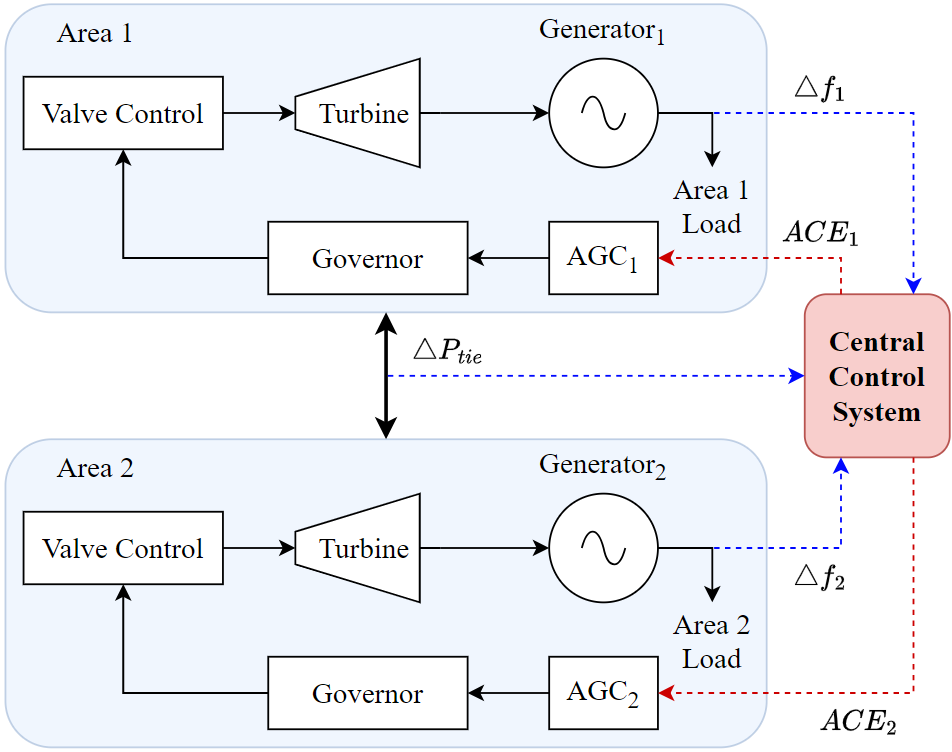} \newline
    {\fontsize{8pt}{10pt}\selectfont (a)}\ \vspace{0.2cm}
    \includegraphics[width=1\columnwidth]{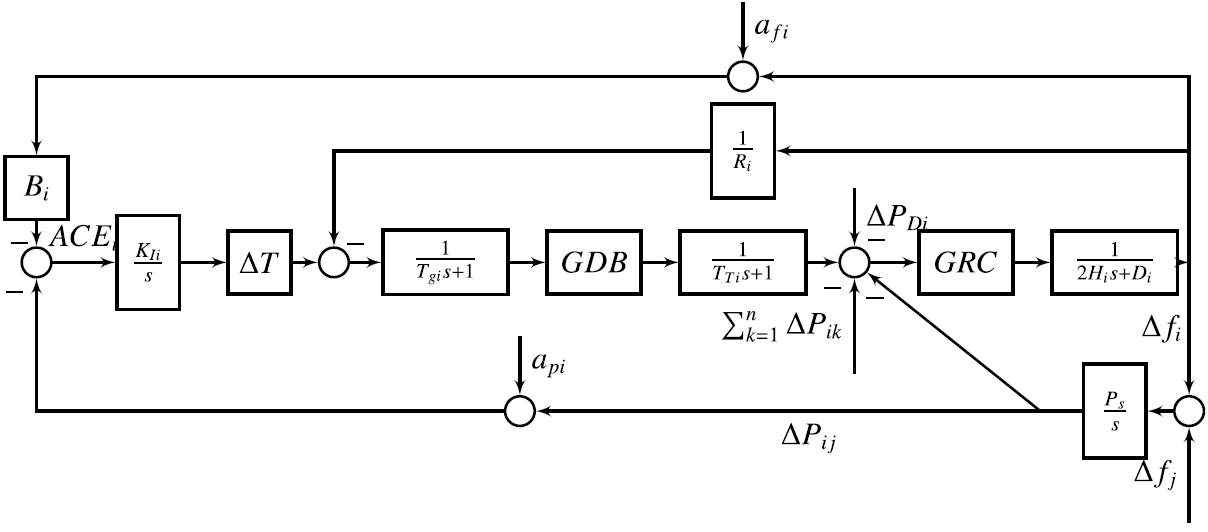} \newline
    {\fontsize{8pt}{10pt}\selectfont (b)} 
    \caption{(a) Schematic of AGC system integrated with a two-area power system, (b) block diagram showing AGC system nonlinearities \cite{ayad2022mitigation}}
    \label{fig:AGC}
\end{figure}

Fig. \ref{fig:AGC} (a) illustrates a typical AGC system controlling a two-area power system, including the nonlinearities inherent in the system shown in Fig. \ref{fig:AGC} (b). 
During disturbances, frequency and tie-line power flow measurements from each area are sent to the control center. The ACE for each area is calculated and relayed to the AGC controller, which adjusts generation accordingly to stabilize the frequency of the interconnected system. The AGC model accounts for several nonlinearities, including GDB, GRC, and time delays \cite{ayad2022mitigation}. These nonlinearities shape the system's response, particularly under attack conditions.
The GDB introduces a threshold where small deviations are ignored by the control system, leading to oscillations around 0.5 Hz. The GRC limits the rate of generation change, preventing rapid adjustments beyond certain thresholds. Time delays due to communication networks also play a significant role in delaying control actions.
These nonlinearities significantly affect the system's behavior and its ability to respond to disturbances, including malicious attacks. The following equations describe the nonlinear AGC model:

\begin{equation}
\Delta P_{mi} - \Delta P_{Di} - \sum_{j=1}^{n} \Delta P_{ij} = 2H_{i} \Delta f_{i} \frac{d \Delta f_{i}}{dt} + D_{i} \Delta f_{i}
\end{equation}
\begin{equation}
\Delta P_{vi} = \Delta P_{mi} + T_{Ti} \Delta P_{mi} \frac{d \Delta P_{mi}}{dt}
\end{equation}

\begin{equation}
x_{i} + \frac{\Delta f_{i}}{R_{i}} = \frac{T_{gi}}{A} \Delta P_{vi} \frac{d \Delta P_{vi}}{dt} + \frac{\Delta P_{vi}}{A}
\end{equation}

\begin{equation}
ACE_{i} =\frac{1}{K_{li} \Delta T} \frac{dx_i}{dt}= \sum_{j=1}^{n} \Delta P_{ij} + B_{i} \Delta f_{i}
\end{equation}

\begin{equation}
\Delta f_{i} - \Delta f_{j} = \frac{1}{P_{s}} \frac{d \Delta P_{ij}}{dt}
\end{equation}

\begin{equation}
B_{i} = \frac{1}{R_{i}} + D_{i}
\end{equation}

\noindent where $\Delta P_{mi}$, $\Delta P_{Di}$, $\Delta P_{vi}$, $H_i$, and $\Delta f_{i}$ are the mechanical power change of the turbine, the applied disturbance, the governor output change, inertia, and the frequency deviation in area $i$, respectively. $\Delta P_{ij}$ is the tie-line power deviation between area $i$ and area $j$. $T_{Ti}$ and $T_{gi}$ are the turbine and governor time constants, respectively. $P_s$ is the synchronization parameter. $D_i$, $B_i$, $x_i$, $K_{li}$, and $x_{i}$ are the load-frequency parameter, frequency bias factor, the output of the integrator, controller parameter, and the parameter of speed regulator, respectively. The turbine output power deviation is limited by the GRC limit and implemented in (1). Similarly, the governor power deviation is limited by the GDB limit and implemented in (2). Furthermore, the time delay effect is shown in (4).

The nonlinearities in real AGC systems, such as GDB, GRC, and time delays, can make it difficult to detect subtle system anomalies. For instance, the GDB masks minor frequency variations, which could potentially be exploited in an attack scenario. Similarly, the GRC constrains the rate of generation change, potentially allowing an attacker to manipulate generation or frequency measurements within permissible limits, thus avoiding detection. Time delays also introduce challenges in the real-time detection of malicious activities, as control actions are not immediately responsive to data changes.

One common cyberattack against AGC systems is the FDIA, where attackers manipulate transmitted frequency or tie-line power flow measurements between generation areas and the control center. These attacks aim to disrupt the system by inducing incorrect ACE calculations, which in turn lead to improper generation regulation.
Further, FDIAs can be designed to be subtle, introducing gradual and controlled data manipulations that avoid triggering traditional bad data detection mechanisms. By carefully crafting false data, attackers can deceive the AGC into making incorrect adjustments to generation, causing frequency deviations, generation mismatches, or even instability without being detected. 
There are various strategies for implementing FDIAs, such as ramp, pulse, or step functions. These attack strategies are tailored to the AGC's response characteristics and can introduce controlled changes over time, which makes the attacks more difficult to detect and mitigate.
To counter these sophisticated threats, advanced detection methods are required that can identify FDIAs on AGC systems considering the system nonlinearities.

\section{ML-based Detection of FDIAs on AGC System}\label{Section:Methodology}
\begin{figure}[t!]
\centering
\includegraphics[width=1\columnwidth]{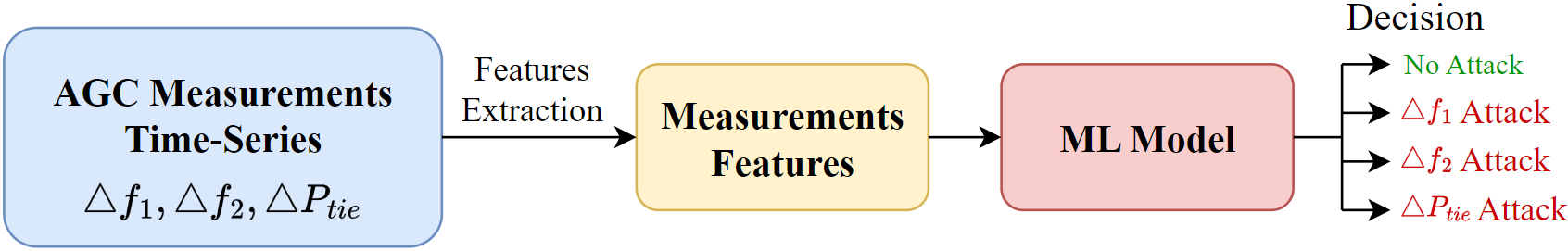}
\caption{Schematic of proposed cyberattack detection and classification scheme}
\label{fig:diagram}
\end{figure}

This section presents a novel ML-based scheme designed to detect FDIAs on the AGC system. The proposed method, illustrated in Fig. \ref{fig:diagram}, utilizes the key AGC measurements—frequency deviations ($\Delta f_1$, $\Delta f_2$ of area 1 and area 2, respectively) and tie-line power flow ($\Delta P_{\text{tie}}$)—to determine whether a disturbance is legitimate or if an attack has been launched on any of these measurements.
The detection process is based on the comparison of the three-dimensional AGC measurements before and after a disturbance. These measurements are processed through a feature extraction phase, where various statistical features are derived. The extracted features are then input into an ML model, which has been trained offline to classify the disturbance as either an authentic event or one of the predefined attack types. Specifically, the model distinguishes between four classes, including the “no-attack” class, which represents normal system operation.

\subsection{Feature Extraction}

We adopt a feature extraction-based approach to transform the raw AGC measurements into distinct features, making them suitable for ML models. The key advantage of this approach is its ability to convert the time-series data into a structured set of features, which are more appropriate for a wide range of ML algorithms that may not be well-suited for raw time-series inputs.
As illustrated in Table \ref{tab:features_ex}, utilized features span several categories, each capturing different aspects of the time-series data. These categories are designed to quantify various properties of the measurements, including statistical moments, energy content, temporal patterns, frequency-domain characteristics, and non-linear dependencies.

The extracted features are designed to capture various aspects of the AGC time-series measurements. Basic statistics, e.g., mean, variance, skewness, summarize the overall distribution and variability of the data. Energy-based features assess the signal's fluctuation and changes over time, while entropy and complexity features highlight irregularities and dynamic complexity. Autocorrelation features capture temporal dependencies, and frequency domain features, derived from FFT coefficients, provide insights into periodic behaviors. Percentile features describe the distribution of values, especially in the tails, and large standard deviation features focus on significant deviations, which can signal abnormal events. Together, these features offer a comprehensive representation of the system's behavior, aiding in the detection of attacks or disturbances.
These diverse features enable the ML model to detect various types of attacks or disturbances in the AGC system. By combining statistical, temporal, frequency-based, and complexity-oriented features, the model can discern between legitimate system behaviors and manipulated measurements, making it robust to a wide range of potential attack scenarios.

After extraction, the number of features is reduced through a filtration process that retains only the most relevant features, further improving the model's performance.
The filtration is done by calculating the p-value for each feature then the Benjamini Hochberg procedure  \cite{thissen2002quick} is leveraged to determine which features to keep based on their importance and relevance for the required classification task.
The procedure controls the False Discovery Rate (FDR), allowing for a balance between selecting relevant features and limiting the inclusion of irrelevant ones. Specifically, each feature's p-value is calculated, and the Benjamini-Hochberg procedure ranks these p-values in ascending order. It then applies a threshold to decide which features to retain based on their significance, ensuring that the most informative features for classifying disturbances or attacks are kept, thus enhancing the model's predictive accuracy while minimizing overfitting.

\begin{table}[t!]
\centering
\caption{Summary of main utilized features}
\label{tab:features_ex}
\begin{tabular}{l|l}
\Xhline{3\arrayrulewidth}
\rule{-3pt}{2ex} 
\textbf{Feature Category} & \textbf{Extracted Features}\\ 
\hline
\rule{-3pt}{2ex}
Basic Statistics     & Mean, Median, Variance, Standard Deviation,\\
 & Skewness, Kurtosis, Maximum, Minimum,\\
 & Sum of Values\\
\hline
\rule{-3pt}{2ex} Energy and Change-     & Absolute Energy, Absolute Sum of Changes,\\
\rule{-3pt}{2ex} Based Features & Mean Absolute Change, Change Quantiles\\
 & (Mean, Standard Deviation)\\
\hline
\rule{-3pt}{2ex} Entropy and     & Sample Entropy,\\
\rule{-3pt}{2ex} Complexity Features &  CID-CE\(^{[1]}\) (Normalized and Non-Normalized)\\
\hline
\rule{-3pt}{2ex} Autocorrelation and     & Aggregated Autocorrelation (Variance,\\ 
\rule{-3pt}{2ex} Nonlinear Features & Max Lag = 32), C3\(^{[2]}\) (lag=1,2,3)\\
\hline
\rule{-3pt}{2ex} Frequency Domain      & FFT Coefficients (orders 0 to 64), FFT \\
\rule{-3pt}{2ex} Features &  Aggregated (Centroid, Variance, Skewness,\\
 &   Kurtosis)\\
\hline
\rule{-3pt}{2ex}Percentiles & 10\(^{th}\), 20\(^{th}\), 30\(^{th}\), 40\(^{th}\), 60\(^{th}\), 70\(^{th}\), 80\(^{th}\), 90\(^{th}\)\\
\hline
\rule{-3pt}{2ex} Large Standard  & Relative Deviation Thresholds of 0.25 and 0.35\\
\rule{-3pt}{2ex} Deviation Features &  
\\ \hline \Xhline{3\arrayrulewidth}
\end{tabular}
\begin{tablenotes}
    \item[1] [1] Complexity Invariant Distance—Complex Exponent.
    \item[2] [2] Third-Order Nonlinearity Statistics.
\end{tablenotes}
\end{table}

\subsection{ML Classifier for FDIA Detection and Manipulated Measurement Identification}

In this paper, we apply an ML classifier to detect and classify FDIAs in the AGC system. Specifically, the task is framed as a 4-class classification problem, where the classes correspond to three types of disturbances (e.g., FDIA on one of the measurements) and a "no-attack" class.
Given the extracted and filtered features \( \mathbf{X} = [\mathbf{x}_1, \mathbf{x}_2, ..., \mathbf{x}_n] \), where \( \mathbf{x}_i \) represents the feature vector derived from the time-series measurements at time \( t_i \), the objective is to classify each instance \( \mathbf{x}_i \) into one of four possible categories. These categories are: 1)  No-attack (\( y = 0 \)), 2)  FDIA on $\Delta f_{1}$ 
    (\( y = 1 \)),
    3) FDIA on 
    $\Delta f_{2}$
    (\( y = 2 \)), and 
    4) FDIA on 
    $\Delta P_{tie}$
    (\( y = 3 \)).
The problem can be then framed as a supervised classification task, where the goal is to learn a function \( f(\mathbf{x}_i) \) that maps the feature vector \( \mathbf{x}_i \) to the correct class label \( y_i \), i.e.,

\begin{equation}
y_i = f(\mathbf{x}_i)
\end{equation}

\noindent  Let \( \mathbf{X} \in \mathbb{R}^{n \times d} \) represent the feature matrix, where \( n \) is the number of instances and \( d \) is the number of selected features per instance. The goal is to train a model that predicts the class \( y_i \) for each \( \mathbf{x}_i \). The function \( f \) can be any standard ML classifier, such as logistic regression, support vector machine (SVM), random forest, or neural networks. In this work, we assume the use of a multi-class classifier.
Given the feature matrix \( \mathbf{X} \), we seek to learn the optimal parameters \( \theta \) of the model such that the prediction \( \hat{y}_i \) is as close as possible to the true label \( y_i \) for each instance. This can be formulated as an optimization problem with the objective

\begin{equation}
\min_{\theta} \sum_{i=1}^{n} \mathcal{L}(f(\mathbf{x}_i, \theta), y_i)
\end{equation}

\noindent where \( \mathcal{L}(\cdot) \) is the loss function, typically the cross-entropy loss for multi-class classification, represented as

\begin{equation}
\mathcal{L}(\hat{y}_i, y_i) = - \sum_{c=0}^{3} \mathbb{1}_{\{y_i = c\}} \log \left( \frac{e^{\hat{y}_i^c}}{\sum_{c'} e^{\hat{y}_i^{c'}}} \right)
\end{equation}

\noindent Here, \( \hat{y}_i^c \) is the predicted score for class \( c \), and \( \mathbb{1}_{\{y_i = c\}} \) is the indicator function that equals 1 if \( y_i = c \), and 0 otherwise. This loss function penalizes incorrect predictions, with the penalty increasing as the predicted probability diverges from the true class.
The classifier's decision rule assigns an instance \( \mathbf{x}_i \) to the class that maximizes the predicted score:

\begin{equation}
\hat{y}_i = \arg\max_c \hat{y}_i^c
\end{equation}

\noindent Thus, after training, the model will output the most likely class label \( \hat{y}_i \) for each instance.
The model is trained on a labeled dataset consisting of both normal and attack samples. During training, the features extracted from the AGC measurements are fed into the classifier, and the model's parameters are updated using optimization techniques such as stochastic gradient descent (SGD) or Adam, which minimize the cross-entropy loss function. Once trained, the model is capable of detecting and classifying new instances of disturbances as they occur in the AGC system.

Several ML classifiers can be applied to this problem.  In this work, we employ a variety of powerful ML models, including random forests, support vector machines, and decision trees \cite{scikit-learn_ensemble}, for FDIA detection and identification in the AGC system. These models are chosen for their robust performance in classification tasks and their ability to handle complex relationships in high-dimensional features, such as those extracted from time-series measurements. 
Together, these models offer a diverse set of approaches for FDIA detection, providing flexibility in tackling different patterns in the AGC measurements. The following section will present the results of applying these models, and others, and evaluate their performance.

\section{Simulation Results}\label{Section:Results}

\subsection{Dataset Generation and Feature Extraction}

\begin{figure}[t!] \centering \includegraphics[width=1\columnwidth]{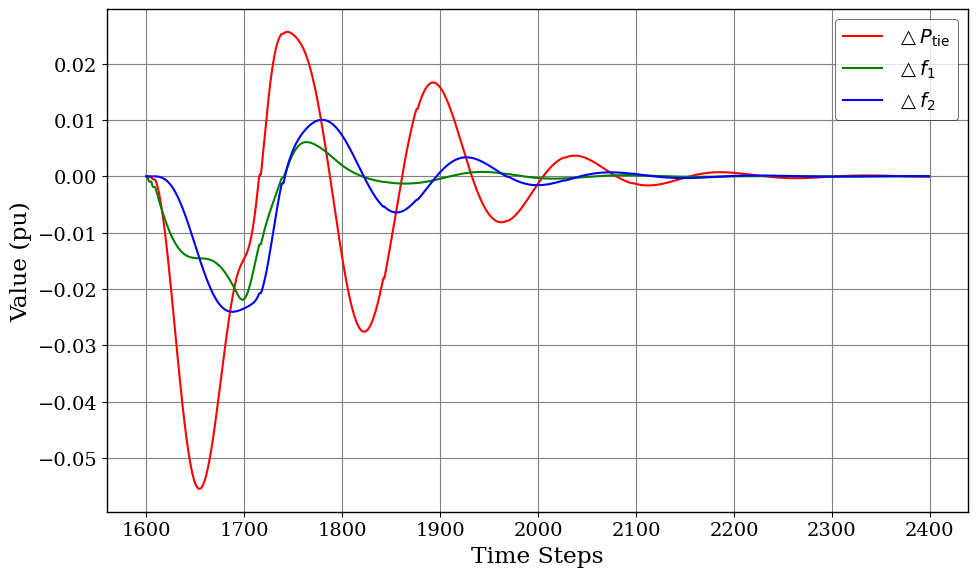} \newline{\fontsize{8pt}{10pt}\selectfont (a)}\ \vspace{0.2cm} \includegraphics[width=1\columnwidth]{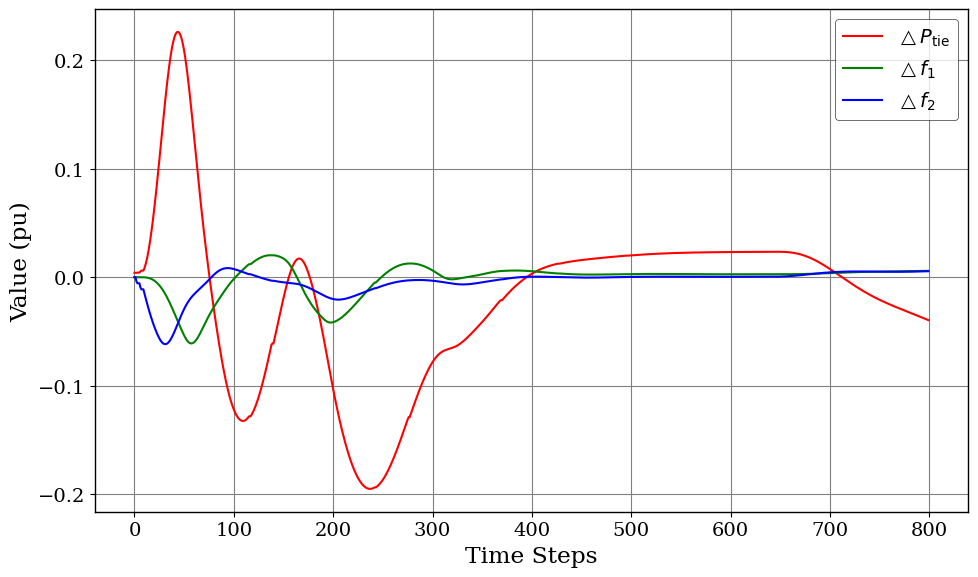} \newline{\fontsize{8pt}{10pt}\selectfont (b)} \caption{(a) A normal disturbance, (b) an FDIA on $\Delta f_{2}$} \label{fig:samples} \end{figure}

 A dataset is generated for a variety of FDIA and normal scenarios using the two-area AGC system from \cite{ayad2022mitigation}. 
 The generated dataset consists of 2400 samples, distributed and labelled as follows: 200 samples correspond to no attack conditions (Class 0), 700 samples correspond to attacks on $\Delta f_{1}$ (Class 1), 700 samples correspond to attacks on $\Delta f_{2}$ (Class 2), 800 samples correspond to attacks on $\Delta P_{\text{tie}}$ (Class 3).  
This imbalance is mainly because cyberattack data samples are more than normal disturbance ones. Cyberattacks can affect different measurements and can take various shapes, e.g., ramp, scaling, and combined attacks. This imbalance poses a challenge for model performance, as it may lead to a bias towards the more prevalent classes. 
Attempts to address this class imbalance through sub-sampling were avoided, as reducing the number of samples would undermine the overall training dataset, particularly in terms of classification performance. 
This dataset of AGC measurements is split into 80\% of the samples for training and  20\% for testing.  
Each sample in the dataset contains readings of the three AGC measurements over an 80-second simulation period for the three measurements in two-area AGC systems.  
 Fig. \ref{fig:samples} illustrates two samples from the dataset.  

Next, a copy of the training dataset is created replacing the raw AGC measurements in each sample with the values of all features explained in Section \ref{Section:Methodology}.  Afterwards, this copy of the training dataset is used to optimize the number of features based on the procedure explained in Section \ref{Section:Methodology}. 
The procedure resulted in retaining 259 features out of the original 300 features.
Using these optimized features, a new training dataset and a new testing dataset are then created out of the original raw-AGC-measurements ones. These two new datasets are used for the remainder of this section.

\subsection{ML Classifier Settings}

Several powerful ML classifiers \cite{scikit-learn_ensemble} are then trained on the training dataset. The main parameters of these classifiers are configures as explained below. 
\subsubsection{Random Forest}
The random forest classifier consists of 500 decision trees and uses the Gini criterion to measure the quality of splits. There is no maximum depth specified for the trees, allowing them to grow as needed. The minimum number of samples required to split an internal node is set to 2.
\subsubsection{Gaussian Naive Bayes}
The Gaussian Naive Bayes classifier is implemented with a variance smoothing parameter of $1\times10^{-9}$. No prior probabilities are specified, meaning the model assumes equal class priors.
\subsubsection{Support Vector Machine}
The SVM classifier employs a linear kernel and utilizes three-fold cross-validation for hyperparameter tuning. The regularization parameter $C$ is optimized over a range of values: $[1\times10^{-4}, 1\times10^{-2}, 1, 1\times10^{2}]$.
\subsubsection{Decision Trees}
The decision tree classifier uses the Gini criterion for measuring split quality. Similar to the random forest, no maximum depth is imposed, and the minimum number of samples required to split an internal node is 2.
\subsubsection{XGBoost}
The XGBoost classifier is configured with a learning rate of 0.025 and employs a multi-class softmax objective function. The model consists of 300 estimators with a maximum tree depth of 5. The minimum child weight is set to 1.2, while the subsample ratio is 0.8, meaning 80\% of the data is used for training each tree. Additionally, 
each tree is built using 80\% of the available features. The gamma parameter, which controls the minimum loss reduction required for further partitioning, is set to 0.066.

\subsection{Results Discussion}

Tables \ref{tab:combined_confusion_matrices} and \ref{tab:classification_results} summarize 
the results of testing the aforementioned models.
The performance evaluation of various ML classifiers highlights key trade-offs between detection accuracy, false alarm rates, and overall classification effectiveness. Given the critical role of AGC in maintaining power system stability, minimizing false positives and false negatives is essential for reliable operation. This section discusses the results from multiple perspectives, including power system operation, cybersecurity, and ML performance using different statistical metrics \cite{saber2023learning}.

From a power system operation perspective, reducing false alarms is crucial to prevent unnecessary control actions that could disrupt system stability. Decision Trees achieve a perfect detection rate for normal disturbances, ensuring that no unnecessary alarms are triggered. However, their ability to detect FDIAs is lower, which means some attacks may go unnoticed. Random Forest and XGBoost offer a better balance, correctly identifying over 95\% of attack cases while maintaining high detection rates for normal conditions. Their overall weighted accuracy remains among the highest, indicating their reliability in distinguishing between attack and no-attack situations.

From a cybersecurity perspective, the primary goal is to detect as many cyberattacks as possible while maintaining a reasonable false alarm rate. Random Forest achieves the highest attack detection rate at 95.28\%, followed closely by XGBoost at 93.45\%. The high recall values, reaching 100\% in some models, indicate that these classifiers successfully detect all actual attack cases, a crucial factor in effective cybersecurity defense. In contrast, Gaussian Naive Bayes struggles with attack detection, making it less suitable for this application.

From an ML perspective, classifier performance is best evaluated using the F1-score, which balances precision and recall. Random Forest and XGBoost achieve the highest F1-scores, 99.88\% in both cases, followed by Decision Trees, 99.98\%. The strong performance of ensemble-based methods, such as Random Forest and XGBoost, suggests that combining multiple weak learners improves robustness in identifying complex attack patterns. K-Nearest Neighbors, performs moderately well, with recall reaching 99.79\% and precision at 98.56\%, detecting 85.71\% of normal disturbances, 84.83\% of attacks on $\Delta f_1$, 88.9\% of attacks on  $\Delta f_2$,   and 93.04\% of attacks on $\Delta P_{\text{tie}}$.
Gaussian Naive Bayes, on the other hand, delivers significantly lower accuracy, likely due to its assumption of feature independence, which does not align well with AGC measurement structures.
%


\begin{table*}[t!]
\centering
\caption{ Confusion matrices (in percentages)}
\label{tab:combined_confusion_matrices}

\begin{tabular}{l|cccc|cccc|cccc}
\Xhline{3\arrayrulewidth}
\multicolumn{1}{c|}{} & \multicolumn{12}{c}{\textbf{Predicted}} \\
\cline{2-13} 
\multicolumn{1}{c|}{\textbf{Actual}} & \multicolumn{4}{c|}{\textbf{Random Forest}} & \multicolumn{4}{c|}{\textbf{Gaussian Naive Bayes}} & \multicolumn{4}{c}{\textbf{SVM}} \\
& No Attack & $\Delta f_1$ & $\Delta f_2$ & $\Delta P_{\text{tie}}$ & No Attack & $\Delta f_1$ & $\Delta f_2$ & $\Delta P_{\text{tie}}$ & No Attack & $\Delta f_1$ & $\Delta f_2$ & $\Delta P_{\text{tie}}$ \\\hline
No Attack & 97.62 & 2.38 & 0 & 0 & 97.62 & 2.38 & 0 & 0 & 95.24 & 0 & 4.76 & 0 \\
$\Delta f_1$ & 0 & 91.03 & 5.52 & 3.45 & 17.93 & 32.41 & 12.41 & 37.25 & 1.38 & 76.55 & 16.55 & 5.52 \\
$\Delta f_2$ & 0 & 2.22 & 95.56 & 2.22 & 20 & 5.19 & 22.22 & 52.59 & 0.74 & 10.37 & 82.96 & 5.93 \\
$\Delta P_{\text{tie}}$ & 0 & 0.63 & 0.63 & 98.74 & 24.68 & 2.53 & 0 & 72.79 & 0 & 5.7 & 1.9 & 92.4 \\
\Xhline{3\arrayrulewidth}
\end{tabular} \\ \textcolor{white}{.}\\

\begin{tabular}{l|cccc|cccc|cccc}
\Xhline{3\arrayrulewidth}
\multicolumn{1}{c|}{} & \multicolumn{12}{c}{\textbf{Predicted}} \\
\cline{2-13} 
\multicolumn{1}{c|}{\textbf{Actual}} & \multicolumn{4}{c|}{\textbf{Decision Trees}} & \multicolumn{4}{c|}{\textbf{XGBoost}} & \multicolumn{4}{c}{\textbf{LSTM} \cite{ayad2022mitigation}} \\
& No Attack & $\Delta f_1$ & $\Delta f_2$ & $\Delta P_{\text{tie}}$ & No Attack & $\Delta f_1$ & $\Delta f_2$ & $\Delta P_{\text{tie}}$ & No Attack & $\Delta f_1$ & $\Delta f_2$ & $\Delta P_{\text{tie}}$ \\\hline
No Attack & 100 & 0 & 0 & 0 & 97.62 & 0 & 2.38 & 0 & 96.67 & 0 & 3.33 & 0 \\
$\Delta f_1$ & 0 & 82.76 & 11.72 & 5.52 & 0 & 88.97 & 6.9 & 4.13 & 0 & 93.25 & 3.37 & 3.38 \\
$\Delta f_2$ & 0.74 & 7.42 & 86.67 & 5.17 & 0 & 4.45 & 93.33 & 2.22 & 0 & 4 & 93.77 & 2.23 \\
$\Delta P_{\text{tie}}$ & 0 & 3.80 & 4.43 & 91.77 & 0 & 1.9 & 0.63 & 97.47 & 0 & 3.89 & 1.56 & 94.55 \\
\Xhline{3\arrayrulewidth}
\end{tabular}
\end{table*}

\begin{table*}[h]
    \centering
    \renewcommand{\arraystretch}{1.2}
    \caption{Performance evaluation metrics (in percentages)}    
    \begin{tabular}{ l|c|c|c|c|c|c }
        \Xhline{3\arrayrulewidth}
        \textbf{Classifier} & \textbf{ Detected FDIAs } & \textbf{ Detected No-Attack Cases } & \textbf{Weighted Accuracy } & \textbf{Precision } & \textbf{Recall } & \textbf{F1-score } \\
        \hline
        Decision Trees & 87.28 & \textbf{100} & 88.3 & \textbf{100} & 99.96 & \textbf{99.98} \\
        Random Forest & \textbf{95.28} & 97.62 & \textbf{95.47} & 99.77 & \textbf{100} & 99.88 \\
        XGBoost & 93.45 & 97.62 & 93.8  & 99.77 & \textbf{100} & 99.88 \\
        LSTM \cite{ayad2022mitigation} & 93.89 & 96.67 & 94.12 & 99.68 & \textbf{100} & 99.84 \\
        SVM & 84.35 & 95.24 & 85.26 & 99.49 & 99.89 & 99.69 \\
        KNN & 89.11 & 85.71 & 88.83 & 98.56 & 99.79 & 99.17 \\
        Gaussian Naive Bayes & 43.85 & 97.62 & 48.33 & 99.51 & 93.90 & 96.58 \\
        \Xhline{3\arrayrulewidth}
    \end{tabular}

    \label{tab:classification_results}
\end{table*}

\subsection{ Comparison with Existing Work and Discussion}
Several models in the proposed approach outperform the results reported in \cite{ayad2022mitigation}, which trained an LSTM model directly AGC time series measurements.
This outcome highlights that the strength of the proposed approach comes not just from the choice of classifier but from the effectiveness of statistical feature extraction in enhancing detection performance. The proposed feature-based methodology allows these models to achieve competitive performance while maintaining interpretability, making them practical solutions for cyberattack detection in AGC systems.
Moreover, existing approaches like \cite{ayad2022mitigation} lack interpretability regarding the features that the LSTM extracts from the AGC measurements. This opacity may hinder its adoption in real AGC systems, unlike the proposed approach in this paper with features that can be easily understood by human operators.

While the proposed approach can accurately detect FDIAs that manipulate the AGC measurements aiming to disrupt the system operation, it cannot detect covert cyberattacks that can be performed with the goal of masking an actual system disturbance from the controller, preventing the AGC system from responding to this disturbance \cite{wang2023deep,saber2024unmasking}. 
This represents an interesting direction for future work. Future work can also include investigating the interpretability, scalability, and real-time performance of the proposed ML-based scheme.

\section{Conclusion}\label{Section:Conclusion}

This paper proposes a new  ML-based approach for detecting FDIAs on AGC systems, taking into account the inherent non-linearities of the system. The proposed approach utilizes a range of statistical and domain-specific features derived from time-series measurements of the AGC system and applies various ML models to detect and classify FDIAs effectively. 
Our results demonstrate that the proposed method can accurately detect FDIAs in AGC systems and identify manipulated measurements while maintaining a low false alarm rate. The approach outperforms existing methods discussed in related works, highlighting its robustness and efficiency in handling the complex dynamics and non-linear behaviors of AGC systems. 
Future work directions have been also discussed.

\ifCLASSOPTIONcaptionsoff
  \newpage
\fi

 \bibliographystyle{IEEEtran} 
 \bibliography{cas-refs.bib}

\end{document}